\documentclass[english,aps,a4paper,prd,superscriptaddress,preprintnumbers,nofootinbib,preprint,floatfix,showpacs]{revtex4}

\usepackage{hyperref}
\usepackage{amsmath}
\usepackage{amssymb}
\usepackage{graphicx}
\usepackage{subfig}
\usepackage{rotating}
\usepackage{color} 
\usepackage{multirow} 
\usepackage[T1]{fontenc} 
\usepackage[utf8]{inputenc}
\usepackage{lmodern}
\makeatletter

\usepackage{booktabs}
\usepackage{mathrsfs}   
\usepackage{slashed}     
\usepackage{url}
\usepackage{multirow}
\usepackage{units}
\usepackage{wasysym}

\def\lnv{lepton number violation }
\def\TrTrOne{ $\mathrm{SU(3)_c \otimes SU(3)_L \otimes U(1)_X}$ }
\usepackage{float}
\newcommand{\bea}{\begin{eqnarray}}
\newcommand{\eea}{\end{eqnarray}}
\def\SM{$\mathrm{SU(3)_c \otimes SU(2)_L \otimes U(1)_Y}$ }
 
 \def\z53{ $\mathbb{Z}_{5} \otimes \mathbb{Z}_{3} $}
\def\ze33{ $\mathbb{Z}^D_{3} \otimes \mathbb{Z}_{3} $}

\newcommand{\sm}{Standard Model }

\makeatother

\usepackage{babel}

\def\vev#1{\left\langle #1\right\rangle}

\def\5{\overline 5}

\newcommand{\diag}{\text{diag}}

\def\vev#1{\left\langle #1\right\rangle}

\def\hbar{\hspace{0pt}\raisebox{1pt}{$-$} \hspace{-7pt} h}

\newcommand{\be}{\begin{equation}}
\newcommand{\ee}{\end{equation}}
\newcommand{\bac}{\beq\begin{array}}
\newcommand{\eac}{\end{array}\eeq}
\newcommand{\ba}{\begin{array}}
\newcommand{\ea}{\end{array}}

\newcommand{\bmat}{\begin{pmatrix}}
\newcommand{\emat}{\end{pmatrix}}

\newcommand{\AddrAHEP}{
  {\it AHEP Group, Instituto de F\'{\i}sica Corpuscular --
    C.S.I.C./Universitat de Val{\`e}ncia \\
    Edificio de Institutos de Paterna,
 C/Catedratico Jos\'e Beltr\'an, 2 E-46980 Paterna (Val\`{e}ncia) - SPAIN}}

\begin{document}

\title{Naturally light neutrinos in \textit{Diracon} model}
\author{Cesar Bonilla} \email{cesar.bonilla@ific.uv.es}
\affiliation{\AddrAHEP}
\author{Jose W.F. Valle} \email{valle@ific.uv.es}
\affiliation{\AddrAHEP}

\begin{abstract}
\vspace{1cm}
  We propose a simple model for Dirac neutrinos where the smallness of
  neutrino mass follows from a parameter $\kappa$ whose absence
  enhances the symmetry of the theory.
  Symmetry breaking is performed in a two--doublet Higgs sector
  supplemented by a gauge singlet scalar, realizing an accidental
  global U(1) symmetry.
  Its spontaneous breaking at the few TeV scale leads to a physical
  Nambu--Goldstone boson -- the \textit{Diracon}, denoted $\mathcal
  {D}$ -- which is restricted by astrophysics and induces invisible
  Higgs decays such as $h\to \mathcal{D} \mathcal {D}$.
  The scheme provides a rich, yet very simple scenario for symmetry
  breaking studies at colliders such as the LHC.

\end{abstract}

\maketitle



\section{Introduction}

Establishing whether neutrinos are their own anti--particles continues
to challenge
experimentalists~\cite{Barabash:2011mf,avignone:2007fu}. Likewise, the
mechanism responsible for generating small neutrino masses remains as
elusive as ever.
It is well--known that, in gauge theories, the detection of
neutrinoless double beta decay would signify that neutrinos
are of Majorana type~\cite{Schechter:1981bd,Duerr:2011zd}.
Although experimental confirmation of the Majorana hypothesis may come
in the not too distant future~\cite{KamLAND-Zen:2016pfg}, so far the
possibility remains that neutrinos can be Dirac particles, despite the
fact that the general theoretical expectation is that they are
Majorana fermions~\cite{Schechter:1980gr} as given, for example, in
Weinberg's dimension five operator~\cite{weinberg:1980bf}.
Moreover, the most widely studied mechanism to account for the
smallness of neutrino masses relative to the charged fermion masses
invokes their Majorana nature, namely, the conventional high--scale
type-I~\cite{gell-mann:1980vs,yanagida:1979,mohapatra:1980ia,Schechter:1980gr,Schechter:1981cv}
or type-II~\cite{Schechter:1980gr,Schechter:1981cv,Lazarides:1980nt}
seesaw mechanism.
The same happens in low--scale variants of the seesaw mechanism,
see~\cite{Boucenna:2014zba} for a review.

Accommodating the possibility of naturally light Dirac neutrinos
constitutes a double challenge.
One approach is to supplement the standard \SM electroweak gauge
structure by using extra flavor symmetries implying a conserved lepton
number, so as to obtain Dirac neutrinos, as suggested
in~\cite{Aranda:2013gga,Kanemura:2016ixx}. Another approach would be
to appeal to the existence of extra dimensions, such as in warped
scenarios~\cite{Chen:2015jta}.
Alternatively, one may extend the gauge group itself, for example, by
using the extended \TrTrOne gauge structure because of its special
features~\cite{Singer:1980sw}. In this case one can obtain both the
lightness as well as the Dirac nature of neutrinos as an
outcome~\cite{Valle:2016kyz}.

In this letter we focus on the possibility of having naturally light
Dirac neutrinos with seesaw--induced masses within the framework of
the simplest four-dimensional \SM gauge structure, without non-Abelian
discrete flavor symmetries. To this end we impose a cyclic
flavor--blind \z53 symmetry in a theory with enlarged symmetry
breaking sector : two Higgs doublets and a singlet, see
Table~\ref{tab:1}.
We find that the resulting model has an accidental spontaneously
broken U(1) symmetry that leads to the seesaw mechanism as well as the
Dirac nature of neutrinos. The smallness of neutrino mass follows from
the smallness of a parameter $\kappa$ whose absence would increase the
symmetry of the electroweak breaking sector, ensuring naturalness in
't Hooft's sense.
We discuss some phenomenological features of the scheme which follow
from the existence of a \textit{Diracon} namely, the Nambu--Goldstone
boson associated to the spontaneous breaking of the global accidental
U(1) symmetry in the scalar sector.

\section{ the model}

\begin{center}

\begin{figure}[h!]
\includegraphics[width=0.2\textwidth]{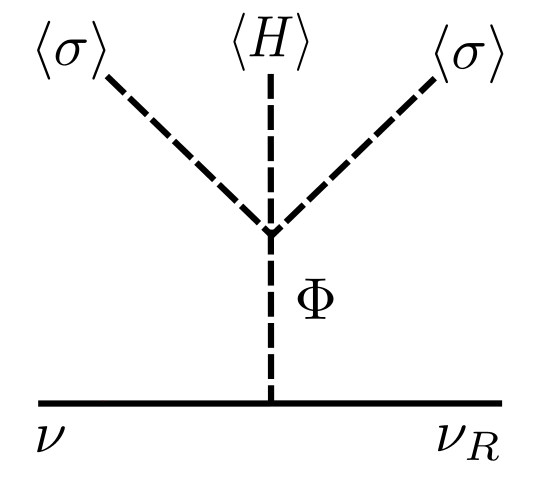}
\caption{ Neutrino mass generation in type-II Dirac seesaw mechanism.\label{Fig:numass}}
\end{figure}
\end{center}

The lepton and scalar boson assignments of the model are summarized in
Table~\ref{tab:1}, where a cyclic \z53 symmetry is assumed, so that
$\omega^5=1$ and $\alpha^3=1$.
The Abelian $\mathbb{Z}_{5}$ symmetry is used to have Dirac neutrinos
in the presence of the additional doublet $\Phi$, forbidding the terms
$\overline{L}\nu_R \tilde{H}$, $\overline{L}\ell_R \tilde{\Phi}$,
$\nu_R \nu_R $ and $\nu_R \nu_R \sigma$. We have used
  two Higgs doublets
 \begin{eqnarray}
  H =\left(\begin{array}{c} 
            h^{+} \\
            H^{0}
           \end{array}\right),\ \
  \Phi =\left(\begin{array}{c}
            \Phi^{0} \\
            \phi^{-}
           \end{array}\right), \notag
 \end{eqnarray}
 with their conjugates defined as usual, $\tilde{H}=i\sigma_2 H^\ast$
 and $\tilde{\Phi}=i\sigma_2 \Phi^\ast$.
On the other hand, the
complementary $\mathbb{Z}_{3}$ symmetry~\cite{Ma:2014qra,Ma:2015mjd}
ensures lepton number conservation also at the non-renormalizable
level, ruling out possible operators of the type $\nu_{R} \nu_{R}
\sigma^3$, $\nu_{R} \nu_{R} \sigma^8$, etc.
\begin{table}[H]
\centering
\begin{tabular}{|c|c|c|c||c|c|c|c|}
\hline
                &    $\overline{L}$ & $\ell_{R}$    & $\nu_{R}$         & $H$     & $\Phi$ &   $\sigma$  \\
\hline
$SU(2)_{L}$     &   ${\bf2}$       & ${\bf1}$     & ${\bf1}$          & ${\bf2}$& ${\bf2}$   &    ${\bf1}$  \\     
\hline
$\mathbb{Z}_{5}$&   $\omega$       & $\omega^4$     & $\omega$          & $1$     & $\omega^3$ &    $\omega$    \\  
\hline
$\mathbb{Z}_{3}$&   $\alpha^2$     & $\alpha$     & $\alpha$          & $1$     & $1$        &    $1$    \\  
\hline
\end{tabular}\caption{Lepton and scalar boson assignments of the model, with 
$\omega^5=1$  and $\alpha^3=1$. }
\label{tab:1}
\end{table}
The gauge-- and \z53--invariant Yukawa Lagrangean for the lepton
sector turns out to be, symbolically,
 \begin{eqnarray}
  \mathcal{L}_{Y} &=& y^{e} \overline{L} e_R H + 
  y^{\nu}  \overline{L} \nu_R \Phi + h.c.
 \end{eqnarray}
 where the first term is the standard one responsible for the charged
 lepton masses, while the second is the one that appears in
 Fig.~\ref{Fig:numass}.
 As illustrated in the figure, the latter induces nonzero neutrino
 masses
 \begin{equation}
  m_{\nu}= \kappa y^\nu \frac{v_\sigma^2  v_H}{m_{\Phi}^2}
 \end{equation}
 where we denote the three vacuum expectation values as
 $v_\sigma\equiv\vev{ \sigma}$, $v_\Phi\equiv\vev{ \Phi}$,
 $v_H\equiv\vev{ H}$, and $\kappa$ is a dimensionless parameter in the
 scalar potential. For simplicity we have omitted generation indices.
 Notice that the smallness of neutrino mass depends not only on the
 Yukawa coupling $y^\nu$ but is also related to the smallness of
 $\kappa$, very much like in the type-II--like seesaw mechanism.
 
\section{  Scalar Sector}

The \SM invariant scalar potential consistent with the global \z53
symmetry is given by~\footnote{This scalar potential is shared
  by other models with Majorana neutrinos. See for example
  Ref.~\cite{Wang:2016vfj}.}
 \begin{eqnarray}
  V &=& -\mu_H^2 H^\dagger H  -\mu_H^2 \Phi^\dagger \Phi - \mu_\sigma^2 \sigma^\dagger \sigma 
        + \lambda_H H^\dagger H H^\dagger H 
        + \lambda_\Phi \Phi^\dagger \Phi \Phi^\dagger \Phi  
        + \lambda_\sigma \sigma^\dagger \sigma \sigma^\dagger \sigma \notag\\
       &+& \lambda_{H\Phi} H^\dagger H \Phi^\dagger \Phi
        + \lambda'_{H\Phi} H^\dagger \Phi \Phi^\dagger H 
        + \lambda_{\sigma H}\sigma^\dagger \sigma H^\dagger H 
        + \lambda_{\sigma \Phi}\sigma^\dagger \sigma \Phi^\dagger \Phi \notag\\
       &+& \kappa \left( \tilde{H}^\dagger \Phi \sigma^2 + h.c.\right)
 \end{eqnarray}
 where, after acquiring vacuum expectation values (vevs) the fields
 are shifted as follows, 
 \begin{eqnarray}
  H^0 = \frac{1}{\sqrt{2}}\left(v_{H}+ R_{H}+i I_{H} \right), \ \
  \sigma = \frac{1}{\sqrt{2}}\left(v_{\sigma}+ R_{\sigma}+i I_{\sigma} \right)\ \ \text{and}\ \
  \Phi^0 = \frac{1}{\sqrt{2}}\left(v_{\Phi}+ R_{\Phi}+i I_{\Phi} \right),
 \end{eqnarray} 
 so the extremum conditions are,
 \begin{eqnarray}\label{minconds}
  \mu_{H}^2&=&\frac{1}{2} \left(2 \lambda _H v_H^2+\lambda_{\sigma  H} v_{\sigma }^2+\lambda_{H\Phi} v_{\Phi }^2-\frac{\kappa  v_{\sigma }^2 v_{\Phi }}{v_H}\right)\notag\\
  \mu_{\Phi}^2&=&\frac{1}{2} \left(\lambda_{H\Phi} v_H^2 +\lambda_{\sigma\Phi} v_{\sigma }^2+2 \lambda _{\Phi } v_{\Phi }^2-\frac{\kappa  v_H v_{\sigma }^2}{v_{\Phi }}\right)\\
  \mu_{\sigma}^2&=&\frac{1}{2} \left(\lambda_{\sigma  H} v_H^2 +\lambda_{\sigma\Phi} v_{\Phi }^2+2 \lambda_\sigma v_{\sigma }^2-2 \kappa  v_H v_{\Phi }\right)\notag
  \end{eqnarray}
  and from these one can derive a ``seesaw--type relation'' amongst the
  vacuum expectation values given as,
\begin{equation}\label{vevss}
v_{\Phi} \approx \kappa v_{H}\left(\frac{1}{\lambda_{H\Phi}\frac{v_{H}^2}{v_{\sigma}^2}
+\lambda_{\sigma\Phi}-2\frac{\mu^2_\Phi}{v_\sigma^2}}\right).
\end{equation}

Notice that $v_\sigma \neq 0$ is required in order to drive $v_\Phi
\neq 0$, see Fig.~\ref{Fig:numass}.  Moreover one sees that the
smallness of $v_{\Phi}$ is directly related to the smallness of
$\kappa$.
In other words, $\Phi$ acquires an ``induced'' vev $v_{\Phi}$ whose
smallness is associated to the symmetry enhancement that results from
the absence of $\kappa$. In this limit there would be a second U(1)
symmetry whose associated Nambu-Golstone boson is the pseudoscalar
$A$, see below.
This means that the ``induced'' vev $v_{\Phi }$ is always very much
suppressed w.r.t. the standard $ v_H$, responsible for generating the
W boson mass. In short the model has a double vev hierarchy
\begin{equation}
  \label{eq:hierarchy}
  v_{\sigma} \gtrsim v_{H} \gg v_{\Phi} .
\end{equation}
The two hierarchies are consistent with the minimization of the
potential. The first is a mild hierarchy, ensuring adequate couplings
for the $Diracon$, while the second one is a strong yet ``natural''
hierarchy, because it is related to the enhanced symmetry which would
result from the absence of $\kappa$ in the Lagrangean, even though, in
practice, it can not be strictly realized, since we need $v_{\Phi}\neq
0$ for a realistic scheme.  \\[-.2cm]

With the information above one can immediately work our the Higgs mass
spectrum.
Out of the ten scalars, eight from the two--doublet structure, plus
two from the extra complex singlet, we are left with seven physical
ones after projecting out the three longitudinal degrees of freedom of
the massive \SM gauge bosons.
These correspond to three physical CP even scalars, one of which is
the 125 GeV state discovered at the
LHC~\cite{Aad:2012gk,Chatrchyan:2012gu,Aad:2015zhl}, two physical CP
odd scalars, and one electrically charged scalar. The mass squared
matrices for the CP-even and CP-odd sectors, in the weak basis
$(H,\sigma,\Phi)$ are given below, 
 \begin{eqnarray}
 M_R^2=\left(
\begin{array}{ccc}
 2\lambda _H v_H^2 +\frac{\kappa  v_{\sigma }^2 v_{\Phi }}{2 v_H}      & \left(\lambda_{\sigma  H} v_H -\kappa  v_{\Phi }\right) v_{\sigma }       & \lambda_{H\Phi} v_H v_{\Phi }  -\frac{\kappa  v_{\sigma }^2}{2} \\
  \left(\lambda_{\sigma  H} v_H -\kappa  v_{\Phi }\right)v_{\sigma }   & 2 \lambda_\sigma v_{\sigma }^2                                            &  \left(\lambda_{\sigma\Phi} v_{\Phi } -\kappa  v_H\right) v_{\sigma } \\
  \lambda_{H\Phi} v_H v_{\Phi }  -\frac{\kappa  v_{\sigma }^2}{2}      & \left(\lambda_{\sigma\Phi} v_{\Phi } -\kappa  v_H\right) v_{\sigma }      & 2 \lambda _{\Phi } v_{\Phi }^2 +\frac{\kappa  v_H v_{\sigma }^2}{2 v_{\Phi }} \\
\end{array}
\right)
\end{eqnarray}
 and
 \begin{eqnarray}
\label{eq:imag}
 M_I^2=\kappa\left(
\begin{array}{ccc}
 \frac{v_{\sigma }^2 v_{\Phi }}{2 v_H} & v_{\sigma } v_{\Phi } & \frac{v_{\sigma }^2}{2} \\
 v_{\sigma } v_{\Phi } & 2 v_H v_{\Phi } & v_H v_{\sigma } \\
 \frac{v_{\sigma }^2}{2} & v_H v_{\sigma } & \frac{v_H v_{\sigma }^2}{2 v_{\Phi }} \\
\end{array}
\right),
\end{eqnarray}
where
$\diag(m_{H_{1}}^2,m_{H_{2}}^2,m_{H_{3}}^2)=\mathcal{O}_{R}M_{R}^2\mathcal{O}_{R}^T$
and $\diag(0,0,m_{A}^2)=\mathcal{O}_{I}M_{I}^2\mathcal{O}_{I}^T$.  \\[-.2cm]

Let us first consider the CP-odd scalars sector, its diagonalization
matrix is given by,
\begin{eqnarray}
\label{OI}
\mathcal{O}_I=\left(
\begin{array}{ccc}
 -\alpha  v_H & 0 & \alpha  v_{\Phi } \\
 -2 \alpha  \beta  v_H v_{\Phi }^2 & \frac{\beta  v_{\sigma }}{\alpha } & -2 \alpha  \beta  v_H^2 v_{\Phi } \\
 \beta  v_{\sigma } v_{\Phi } & 2 \beta  v_H v_{\Phi } & \beta  v_H v_{\sigma } \\
\end{array}
\right)
\end{eqnarray}
where
\begin{equation}
\label{alpbet}
\alpha=\frac{1}{\sqrt{v_H^2+v_{\Phi }^2}}\ \ \text{and}\ \ 
\beta=\frac{1}{\sqrt{v_H^2 \left(v_{\sigma }^2+4 v_{\Phi }^2\right)
+v_{\sigma }^2 v_{\Phi }^2}}.
\end{equation}
Hence one finds that the mass--eigenstate profiles are given by
\begin{eqnarray}
\label{eq:profiles}
G^0&=&\alpha(- v_H I_{H} +  v_{\Phi } I_{\Phi} )\notag\\
\mathcal{D}&=&\alpha  \beta (-2   v_H v_{\Phi }^2 I_{H}+ \frac{v_{\sigma }}{\alpha^2} I_{\sigma}-2 v_H^2 v_{\Phi } I_{\Phi }) \\
A&=& \beta \left(v_{\sigma } v_{\Phi } I_{H}+ 2  v_H v_{\Phi } I_{\sigma}+  v_H v_{\sigma } I_{\Phi}\right).\notag
\end{eqnarray}
One sees that the projective nature of Eq.~(\ref{eq:imag})
  (two--dimensional null space) clearly implies two massless states
  whose profiles follow just from symmetry reasons.
  Due to the smallness of $v_\Phi$, the main components of $G^0$,
  $\mathcal{D}$ and $A$ are the imaginary parts of the $SU(2)_L$ Higgs
  doublet $H$, the singlet $\sigma$ and the doublet $\Phi$,
  respectively.
  Indeed the first massless CP--odd eigenvector $G^0$ pointing mainly
  along $H$ corresponds to the unphysical longitudinal mode of the Z
  boson. The second massless state $\mathcal{D}$ is mainly singlet and
  we call it the \textit{Diracon}, i.e. the physical Nambu-Golstone
  boson associated to the accidental U(1) symmetry. It is the analogue
  of the Majoron present in the ``123'' type-II seesaw scheme
  of~\cite{Schechter:1981cv}, and is associated with the type-II Dirac
  neutrino seesaw mechanism illustrated in Fig.~\ref{Fig:numass}.
  On the other hand the massive pseudoscalar state $A$ pointing mainly
  along the weak isodoublet direction has mass
\begin{equation}
m_A^2= \frac{\kappa  \left(v_H^2 \left(v_{\sigma }^2
+4 v_{\Phi }^2\right)+v_{\sigma }^2 v_{\Phi }^2\right)}{2 v_H v_{\Phi }},
\end{equation}
which would vanish in the (unphysical) limit $\kappa = v_{\sigma} =
v_{\Phi } = 0$.\\[-.2cm]

Turning now to the charged sector we have, in the basis
$\left(h^{\pm}, \phi^{\pm} \right)$, the following mass squared matrix,
 \begin{eqnarray}
 M_{H^\pm}^2=\left(
\begin{array}{cc}
 \left(\lambda'_{H\Phi} v_{\Phi } + \frac{\kappa  v_{\sigma }^2}{v_H}\right) v_{\Phi }& -\lambda'_{H\Phi} v_H v_{\Phi }-\kappa  v_{\sigma }^2  \\
 - \lambda'_{H\Phi} v_H v_{\Phi }-\kappa  v_{\sigma }^2  &  \left(\lambda'_{H\Phi} v_H +\frac{\kappa  v_{\sigma }^2}{v_{\Phi }}\right)v_H \\
\end{array}
\right)
\end{eqnarray}
whose eigenstates are the longitudinal W--boson and a physical
state $H^{\pm}$ of (squared) mass
\begin{equation}
m_{H^{\pm}}^2= \left(v_H^2+v_{\Phi }^2\right) \left(\lambda'_{H\Phi}+\frac{\kappa  v_{\sigma }^2}{v_H v_{\Phi }}\right).
\end{equation}

Notice that, taking into account the smallness of the neutrino mass,
i.e. $\kappa\ll1$, and using Eq.~\ref{vevss} one finds that the Higgs
mass spectrum further simplifies to,
\begin{eqnarray}
M_R^2\approx
\left(
\begin{array}{ccc}
 2 \lambda _H v_H^2  & \lambda_{\sigma  H} v_H v_{\sigma }  & 0 \\
\lambda_{\sigma  H} v_H v_{\sigma }  & 2 \lambda_\sigma v_{\sigma }^2  & 0 \\
 0 & 0 & \frac{\lambda_{H\Phi} v_{H}^2 +\lambda_{\sigma\Phi}v_{\sigma }^2}{2} \\
\end{array}
\right)
\end{eqnarray}

\begin{equation}
\label{mAap}
 m_A^2\approx \frac{\lambda_{H\Phi} v_{H}^2 +\lambda_{\sigma\Phi}v_{\sigma }^2}{2}
\end{equation}

\begin{equation}
\label{mCHap}
m_{H^{\pm}}^2\approx \lambda_{\sigma\Phi}\frac{v_{\sigma}^2}{2}+ \left(\lambda_{H\Phi} + \lambda'_{H\Phi}\right) \frac{v_{H}^2}{2}
\end{equation}

Comparing the CP--even and CP--odd sectors it follows that
$m_{H_3}\approx m_{A}$. Hence by using Eq.~(\ref{mCHap}) and
Eq.~(\ref{mAap}) we find that the following mass relation holds,
\begin{equation}
 m_{H^{\pm}}^2-m_{A}^2\approx \lambda'_{H\Phi} \frac{v_{H}^2}{2}. \notag
\end{equation}

\section{Phenomenological considerations }

\begin{center}
\begin{figure}[h!]
\includegraphics[width=0.35\textwidth]{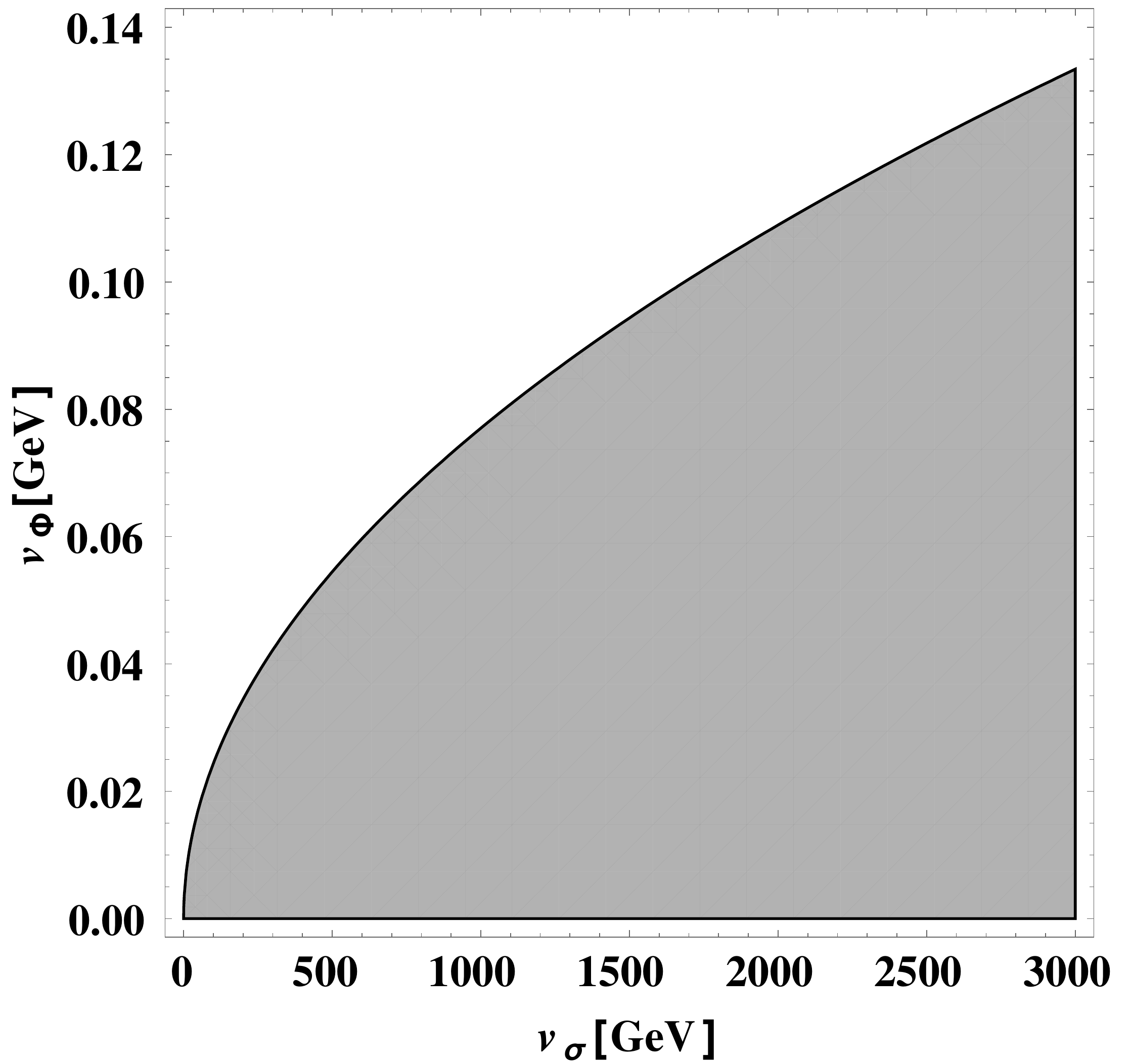}
\caption{The shaded region is allowed by stellar energy loss limits
    .\label{Fig:gDee}}
\end{figure}
\end{center}

The above model of electroweak breaking is rather similar to the inert
doublet model~\cite{branco2012theory}, implying the absence of
tree--level flavor-changing neutral currents. There are, however,
important new features.
A noticeable difference of this model when compared with various
variants of two--Higgs doublet models is the existence of the
accidental U(1) symmetry.
This global symmetry is spontaneously broken by the vev of $\sigma$
implying the existence of a corresponding Nambu--Goldstone boson given
in Eq.~(\ref{eq:profiles}). Its couplings to neutrinos can be easily
obtained from Noether's theorem. Likewise one can determine its
coupling to charged leptons, for instance electrons. The latter would
lead to excessive stellar cooling through the Compton--like process
$\gamma+e \to e +\mathcal {D} $~\cite{Viaux:2013lha}, unless
\begin{equation}
      |g_{ee\mathcal {D}}| = 
\left|(\mathcal{O}_{I})_{21} \frac{m_{e}}{v_{H}}\right|\lesssim 10^{-13}
\end{equation}
hence, using Eq.~(\ref{OI}), one finds
\begin{equation}
\label{APbound}
 2 \alpha  \beta v_{\Phi }^2 \lesssim\frac{10^{-13}}{m_{e}}
\end{equation}
  where $\alpha$ and $\beta$ are given
  in Eq.~(\ref{alpbet}). Taking into account that
  $v_H=\sqrt{v_{SM}^2-v_{\Phi}^2}$ (where $v_{SM}=246$~GeV), one
  writes Eq.~(\ref{APbound}) only in terms of $v_{\sigma }$ and
  $v_{\Phi }$. The allowed region of these vevs is delimited by the
  bound on $g_{ee\mathcal {D}}$ as illustrated in
  Fig.~\ref{Fig:gDee}.  
The shaded area is the region allowed by stellar energy loss limits.

Let us now comment on boundedness conditions and vacuum
stability. Taking $\kappa\ll1$, one can see that the copositivity
criterium applies so one can easily obey the relevant
conditions~\cite{Kannike:2012pe}. Concerning stability and
perturbativity, we finds extented regions of consistency as a result
of the presence of the extra scalar boson
states~\cite{Bonilla:2015kna,Bonilla:2015eha}.
Moreover, in the limit $m_{H_{3}}\sim m_A \sim m_{H^{\pm}}$, it is
well--known that the oblique S,T,U parameters are well under control,
so that these precision observables does not pose great problems
either~\cite{Grimus:2007if,Grimus:2008nb}.

Concerning collider phenomenology, notice that the physical
  Nambu--Goldstone boson induces invisible Higgs boson decays $h\to
  \mathcal{D} \mathcal {D}$.
  These decays are rather analogous to the invisible CP--even Higgs
  decays into \textit{Majorons} which are present in Majorana neutrino
  schemes, such as the ``123'' seesaw~\cite{Schechter:1981cv} with
  spontaneous \lnv~\cite{Diaz:1997xv}. Likewise, one has the new
  pseudoscalar decays $A\to h \mathcal {D}$ and $A\to \mathcal{D}
  \mathcal {D} \mathcal {D}$ in addition to the \sm decay modes. For
  charged scalars, there are also new decay channels into leptons,
  i.e. $H^{\pm}\to \ell^{\pm}\nu_{R}$, which should be taken into
  account in search analyses~\cite{Aad:2015typ}.
    Future experimental searches at the LHC should probe the theory in
    a rather significant way within a relatively wide region of
    parameters.

\section{Summary and conclusions }

We presented a very simple model where neutrinos are Dirac fermions
and their mass can naturally arise from TeV--scale physics, associated
to a small parameter $\kappa$ whose absence would enhance the symmetry
of the theory.
This is realized in an enlarged scalar sector consisting of two
doublets and a singlet Higgs carrying an accidentally conserved global
U(1) charge.
Its spontaneous violation leads to the existence of a physical
Nambu--Goldstone boson which is restricted by astrophysics.
Let us mention that the presence of gravity could induce masses for
neutrinos~\cite{Dvali:2016uhn} and/or the
Diracon~\cite{coleman:1988tj} breaking the Abelian discrete 
symmetries. This breaking may, however, be avoided if the latter are part of a
gauge discrete symmetry~\cite{krauss:1989zc}.
We have discussed the symmetry structure of the model, the connection
to neutrino mass generation, and indicated how it provides new
collider signatures induced by invisible Higgs boson decays.
In summary, the model provides an interesting scheme for neutrino mass
generation. Its scalar sector constitutes an interesting alternative
for electroweak symmetry breaking studies, both theoretically and
experimentally.  Its simplicity, its close connection to neutrino
masses and the presence of an accidental global U(1) symmetry give it
unique features. Details as well as additional phenomenological
features will be discussed elsewhere.\\[-1cm]

\section*{Acknowledgements}

We thank Thomas Neder and Eduardo Peinado for useful comments. This
work is supported by the Spanish grants FPA2011-2297, FPA2014-58183-P,
Multidark CSD2009-00064, SEV-2014-0398 (MINECO) and
PROMETEOII/2014/084 (GVA).

\end{document}